\documentclass[conference]{IEEEtran}

\usepackage[a4paper,%
            left=1.2 cm,right=1.2 cm,top=1.8 cm,bottom= 3.7 cm,%
            ]{geometry}

\IEEEoverridecommandlockouts 
\usepackage{macros}

\usepackage{amsmath}
\usepackage{blindtext, graphicx}
\usepackage{algorithm}

\usepackage{bbm}
\usepackage{soul}

\usepackage{algpseudocode}
\usepackage{pifont}
\usepackage{tikz}
\usepackage{amssymb}
\usepackage{multirow}

\usepackage{upgreek}

\usepackage{mathtools}

\usepackage{textcomp}
\usepackage{hyperref}
\usepackage{lipsum}

\newcommand\copyrighttext{%
  \footnotesize \textcopyright 2018 IEEE. Personal use is permitted, but republication/redistribution requires IEEE permission. This paper is accepted at IEEE SmartGridComm 2018
  DOI:  10.1109/SmartGridComm.2018.8587514 }
\newcommand\copyrightnotice{%
\begin{tikzpicture}[remember picture,overlay]
\node[anchor=south,yshift=10pt] at (current page.south) {\fbox{\parbox{\dimexpr\textwidth-\fboxsep-\fboxrule\relax}{\copyrighttext}}};
\end{tikzpicture}%
}

\ifCLASSINFOpdf
 
\else
  
\fi

\begin{document}

\title{ A unified decision making framework for supply and demand management in microgrid networks}

\author{
	
	\IEEEauthorblockN{D. Raghuram Bharadwaj$^\dagger$}
	\IEEEauthorblockA{Department of CSA,\\
		Indian Institute of Science, 
		Bangalore, India \\
	Email: raghub@iisc.ac.in} \and
\IEEEauthorblockN{D. Sai Koti Reddy$^\dagger$,\\
Krishnasuri Narayanam}
\IEEEauthorblockA{IBM Research - India\\
Email: saikotireddy@in.ibm.com \\
Email: knaraya3@in.ibm.com}
\and
\IEEEauthorblockN{Shalabh Bhatnagar}
\IEEEauthorblockA{Department of CSA and RBCCPS\\
Indian Institute of Science, 
Bangalore, India \\
Email: shalabh@iisc.ac.in}
\thanks{$^\dagger$ Both of these authors are joint first authors, in this paper}
}

\maketitle
\copyrightnotice

\begin{abstract}
This paper considers two important problems - on the supply-side and demand-side respectively and studies both in a unified framework. On the supply side, we study the problem of energy sharing among microgrids with the goal of maximizing profit obtained from selling power while at the same time not deviating much from the customer demand. On the other hand, under shortage of power, this problem becomes one of deciding the amount of power to be bought with dynamically varying prices. On the demand side, we consider the problem of optimally scheduling the time-adjustable demand - i.e., of loads with flexible time windows in which they can be scheduled. While previous works have treated these two problems in isolation, we combine these problems together and provide a unified Markov decision process (MDP) framework for these problems. We then apply the Q-learning algorithm, a popular model-free reinforcement learning technique, to obtain the optimal policy. Through simulations, we show that the policy obtained by solving our MDP model provides more profit to the microgrids. 

\end{abstract}

\section{Introduction}

A microgrid is a networked group of distributed energy sources with the goal of
generating, converting and storing energy. 
While the main power stations are highly connected, microgrids with local power generation, storage
and conversion capabilities, act locally or share power with a few neighboring microgrid nodes \cite{farhangi2010path}.
This scenario is being envisaged as an important alternative to the conventional scheme with
large power stations transmitting energy over long distances.

In order to take full advantage of the modularity and flexibility of microgrid technologies, smart control mechanisms are required to manage and coordinate these distributed energy systems so as to minimize the costs of energy production, conversion and storage, without jeopardizing the central smart grid stability.
 Augmenting microgrid with smart controls however involves addressing many problems. In this paper, we address two  problems. (i) Supply-side management (SSM) problem: energy sharing among  microgrids under stochastic supply and demand along with  optimal battery scheduling of each microgrid and (ii) Demand-side management (DSM) problem: efficiently scheduling the time adjustable demand from smart appliances in a smart home environment along with non-adjustable demand. Our goal here is to maximize profit earned by
microgrids from selling excess energy while maintaining a low
gap between demand and supply. 
 %reduce the energy demand and supply deficit in the long-run.
 We address these learning and scheduling problems by modeling them in the framework of Markov decision process (MDP) \cite{puterman2014markov}.
 
A simple example which explains the specific problem that we are trying to solve on the supply side using our proposed framework is the following. Consider three microgrids MG-1, MG-2, and MG-3 with their respective forecasted demand and supply profiles over two time intervals as in Table \ref{tab:MGprofile}. Here, supply denotes the power available to the microgrid from its renewable energy sources (though, accurate prediction of the supply from renewable energy sources is a challenge). Let us assume for this example, that the microgrids do not buy power from the central main grid to which they are connected. Let the price of the power (per unit) in time interval 2 be higher than the price in time interval 1 (this information is not known to the microgrids a priori). Below are three possible power sharing scenarios, between these microgrids. 

\begin{table}[H]
\begin{center}
\caption {Sample forecasted demand \& supply profiles} \label{tab:MGprofile}
\begin{tabular}{|c|c|c|c|c|}\hline 
\multirow{2}{*}{\bf  Microgrid \#} & \multicolumn{2}{|c|}{\bf  Interval 1 } & \multicolumn{2}{|c|}{\bf  Interval 2 } \\
\cline{2-5}
& demand & supply & demand & supply \\\hline
MG-1 & 1 & 2 & 1 & 0 \\ \hline
MG-2 & 1 & 0 & 1 & 1 \\ \hline
MG-3 & 1 & 2 & 1 & 1 \\ \hline
\end{tabular}
\end{center}
\end{table}

 \begin{enumerate}[label=\textbf{Scenario  \arabic*:},itemindent=*,leftmargin=0em]
\item Power sharing is not allowed between microgrids. Then there is power deficiency in microgrid MG-2 during the time interval 1 and in the microgrid MG-1 during the time interval 2.
\item Power sharing is allowed between microgrids, MG-2 buys power from MG-1 in interval 1, MG-3 stores it's excess power in the battery in interval 1, and MG-1 buys power from MG-3 in interval 2. 
\item Power sharing is allowed between microgrids, MG-2 buys power from MG-3 in interval 1, and MG-1 stores it's excess power in the battery in interval 1 for consumption in interval 2.
\end{enumerate}

Scenario 2 addresses the power deficiency issue in Scenario 1. However a more intelligent way of handling the power deficiency for MG-1 is possible in Scenario 3. In the Scenario 2, the profit obtained by MG-1 (which is negative as it sells the power when the price is low and buys it when the price is high) is less than the profit obtained by it in the Scenario 3 (which is zero). Our objective, therefore in this framework, is to obtain an optimal power buying and selling policy for microgrids, so as to maximize the overall profits , in the presence of renewable energy supply sources and dynamically varying power prices.

An assumption we make in this work is that, in any microgrid, when supply meets the demand in a given interval, it is to be understood that the demand represents the actual power consumption by the consumers, plus the electric power transmission and distribution losses. We however do not explicitly try to minimize the power transmission and distribution losses. Our framework inherently tries to minimize these power losses, by drawing power preferably from nearby peer microgrids as opposed to far-away microgrids. 

\subsection{Supply-side management (SSM) problem}
\begin{figure}[thpb]
      \centering
      \includegraphics[scale=0.31]{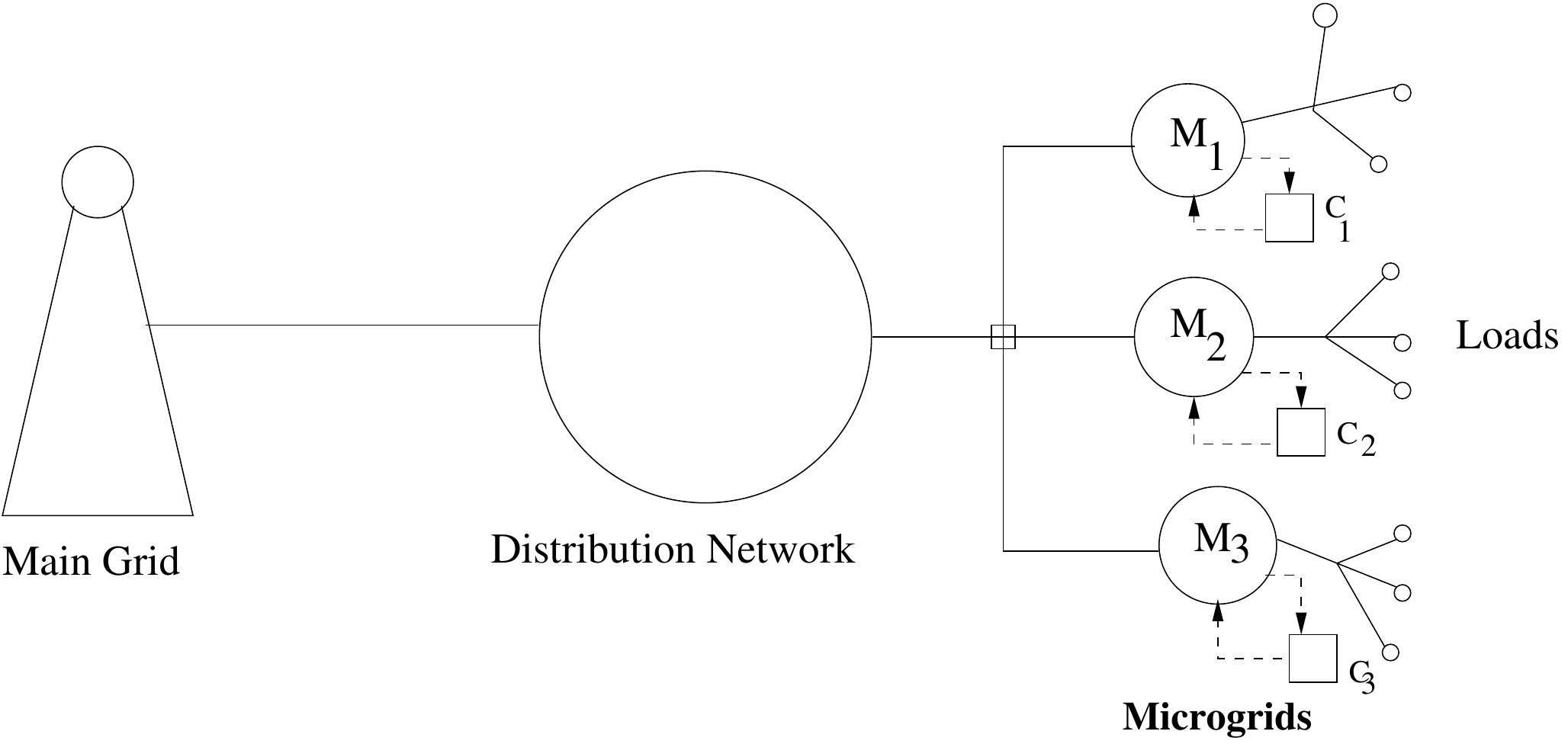}
      \caption{Cooperative Energy Exchange Model}
      \label{gridmodel}
\end{figure}

%Supply-side management (SSM)\cite{} deals with developing techniques to  generate, transmit and distribute energy efficiently at the supply-side.
 Cooperative energy exchange among microgrids is a popular technique in SSM for efficient energy distribution.  Local energy sharing/exchange between microgrids has the
following advantages:
(a) it can significantly reduce power wastage that would
otherwise result over long-distance transmission lines, and (b) it
helps satisfy demand and reduce reliance on the main/central grid. Figure~\ref{gridmodel} shows a cooperative energy exchange model with multiple microgrids
(on the distribution side of the network) that can cater to their individual
local loads. Each microgrid controls its local sub-network through its controller (labeled
$\mbox{C}_1$, $\mbox{C}_2$ etc.) that mainly has access to its local state information.

In classical power grids, system level optimization is done based on a centralized
objective function, where as a
microgrid network has heterogeneous nature right from the manner in which electricity
is generated such as from wind turbines, solar farms and diesel generators
to energy storage devices such as batteries and capacitors.
 Because of this heterogeneity and the fact that energy can be shared between microgrids depending
on requirements, one needs to consider distributed techniques 
%such as multi-agent reinforcement learning or game theory
 to control and optimize a smart grid system
with a distribution network catering to multiple microgrids.

\textbf{Related work :} 
The literature considering the energy exchange among the microgrids is vast.  A survey on game theoretic approaches for microgrids is considered in \cite{saad2012game}. This survey examines both cooperative energy sharing models as well as non-cooperative game models for distributed control of microgrids assuming that the system model is known. Energy sharing among the microgirds is studied in \cite{liu2015energy} with the objective of minimizing the energy bills for the microgrids. This work is later extended in \cite{liu2017energy} to consider the price-based demand response. \cite{mao2014multiagent,cintuglu2015real,dou2016mas} and \cite{logenthiran2015intelligent} consider multi-agent systems for energy trading and control of microgrids under various objectives and formulations.
However, most of the existing literature assumes that the underlying distributions for both supply and demand are known. But this doesn't hold good especially in the context of microgrids with renewable energy sources as there is significant randomness in the amount of energy generated. Since  models for energy dynamics are very unreliable \cite{zamora2010controls} due to randomness at various stages, one needs to use model-free and data-driven algorithms to address these problems. Because of their model-free nature, Reinforcement Learning (RL) \cite{sutton1998reinforcement} approaches that are primarily data-driven control techniques play a significant role in solving these problems. The first step in this direction is to formulate the problem in the Markov Decision Process (MDP) framework. In section \ref{sec:model}, we provide the details of our proposed model for the energy trading problem among the microgrids in order to maximize profits, adhering to the average-cost MDP framework \cite{avgcost}.

\subsection{Demand-side management (DSM) problem}
Load shifting is a popular technique used in demand-side management (DSM) \cite{DTU2010}. It involves moving the consumption of load to different times within an hour,  a day, or  a week. It does not result in reduction in the net quantity of energy consumed, but simply involves changing the time when the energy is consumed. Load shifting facilitates the customer in reducing the energy consumption cost and at the same time it helps the smart grid in managing the peak load.

With increased use of smart appliances and smart home environments, the concept of load shifting is becoming increasingly popular for the smart grid as the demand from smart appliances is time adjustable in general. One or more of these smart appliances collectively achieve some activity in the smart home environment, called ADL (activity of daily living). It is possible to monitor and identify the ADLs in smart home environments \cite{GPG2016}. % When an ADL is active, the smart appliances associated with that ADL are switched on to perform the activity defined by the ADL thus adding load to the smart grid.
  With the help of smart home technology, it is possible to find the amount of load each ADL puts on the grid, and also the allowed time window during which the ADL would perform the activity (e.g., scheduling a  washing machine for an hour to clean the clothes anytime between 3PM to 6PM). 
% If the time window for the ADL lets the smart grid have more than one possible way of scheduling the load, it is considered  flexible. On the other hand, if the time window for the ADL lets the smart grid have exactly one possible way of scheduling the load, then it is  considered  non-flexible.
 The demand from  ADLs need not be met during a fixed time period, instead it could be met during any time period within a flexible time window. With the help of the advanced metering infrastructure (AMI) that provides a two-way communication between the utility and customers, it is possible to make a decision on when to schedule the ADL demand at the smart grid and convey the same to the customer's smart meter. 

There is  regular demand that needs to be met at fixed time periods, apart from the ADL related demand associated with any customer. This regular demand of a smart home will be called non-ADL demand in the rest of the paper. Similarly, the demand from  ADL of the smart home will be called ADL demand.

There is prior work around scheduling the ADL-demand using the load shifting technique for handling  peak load scenarios \cite{CL2014}. However, the authors make the unrealistic assumption of precisely knowing the supply profile while doing such a scheduling of the ADL-demand. In this paper, we propose scheduling of ADL-demand using the load shifting technique with uncertainty in the supply profile generated (e.g., renewable energy sources like solar or wind being the primary sources of power generation).

\textbf{Our main contributions:}\\
\begin{inparaenum}[\bfseries (i)]
\item To the best of our knowledge, we are the first  to integrate both the demand-side and supply-side management problems of a network of microgrids in a unified Markov decision process framework. We apply Reinforcement Learning (RL) algorithms which do not require knowledge of the underlying system model to address these problems. Our algorithms are easy to implement.\\
\item We perform for the first time, optimal scheduling of ADL demand when both demand and power generation are stochastic in nature. Even though this is the most natural scenario, it had not been studied previously. \\    
\end{inparaenum}
The rest of the paper is organized as follows. In section \ref{sec:model}, we discuss in detail about the problem formulation using the MDP framework. We present  in section \ref{sec:algo} the Q-learning algorithm. In section \ref{sec:experiments}, we present simulation experiments along with other algorithms for comparison. Finally, in section \ref{sec:conclusion}, we provide the concluding remarks.

\IEEEpeerreviewmaketitle

\section{Problem formulation and the mdp model} \label{sec:model}
%Microgrids comprise of the distributed small scale power generating sources, mainly renewable energy sources (mainly wind and solar ) and  are also equipped with storage devises.
We consider $N$ microgrids denoted by $1,\ldots,N $ which are inter-connected through the central electric grid distribution network. Each microgrid comprises of the distributed small scale renewable power generation sources that are equipped with energy storage devices. We divide a day into $t$ time units during which the decisions about power allocation are made. At every time instant $t$ of a day, the $i^{th}$ microgrid controller $C_i$ has access to the following information:
\begin{enumerate}[label=(\alph*)]
\item Total renewable energy ($r_t^i$) generated from it's energy sources.
\item Price per unit energy ($p_{t}$) decided by the main grid \\ at time $t$. 
% (super script $i$ used to refer to microgrid $i$).
\item Accumulated non-ADL demand ($d_t^i$) from each load. 
\item Set of all ADL jobs ($J_{t}^{i}$). $J_{t}^{i}$ has the form $\{\gamma_{1}^{i},\ldots,\gamma_{n}^{i}\}$, where the $j^{th}$ ADL job $\gamma_{j}^{i} = (a_{j}^{i}, f_{j}^{i})$. Here, $a_{j}^{i}$ represents the number of units of energy required to finish the job, and  $f_{j}^{i}$ represents the number of future time instants remaining (after the current time instant $t$) by when the controller $C_i$ can schedule the job $\gamma_j^i$ without incurring a penalty.
\item Let $B_{i}$ represent the maximum battery capacity of the microgrid $i$ and $b_{t}^{i}$ the amount of power available in the battery at time $t$. Here $0 \le b_{t}^{i} \le B_{i}$ for any time instant $t$.  
\end{enumerate} 
%In this paper, we consider the cooperative energy exchange model under which microgrids can share energy among themselves. 
From the above available information, microgrid controller  $C_i$ at every time step $t$ has to decide on the following choices: 
\begin{enumerate}[label=(\alph*)]
\item  Amount of energy it needs to buy (sell) from (to) the main grid.
\item Amount of energy it needs to buy (sell) from (to) the neighboring microgrids.
\item Amount of energy it needs to store (retrieve) into (from) its storage device.
\item The subset of ADL jobs it needs to schedule. 
\end{enumerate}
Both the demand and energy generated at each microgrid are uncertain due to the random nature of loads ($d_t^i$ and $J_t^i$) and the amount of renewable energy generation ($r_t^i$) in time slot $t$. Therefore, this problem falls in the realm of Markov Decision Process (MDP). In the next subsection we provide the details of our MDP model.

%***************************************************************

\tikzstyle{decision} = [diamond, draw, fill=gray!30, 
    text width= 4.8 em, text badly centered, node distance=3cm, inner sep=0pt]
\tikzstyle{block} = [rectangle, draw, fill=blue!20, 
    text width= 6.5 em, text centered, rounded corners, minimum height=2.5em]
\tikzstyle{line} = [draw, -latex']
\tikzstyle{cloud} = [draw, ellipse,fill=red!20, node distance=3cm,
    minimum height=2em]
 \begin{figure}
 \centering
  
 \begin{tikzpicture}[node distance = 3cm, auto] 
    % Place nodes
 %  \node [block] (RL) {RL Agent};
    \node [block, text width = 7 em] (state) { \footnotesize Based on the state $s_{t}$, RL agent takes actions $u_{t}$ and $v_{t}$.};
    %\node [block, right of = state, node distance = 3.5 cm ] (action) {\footnotesize Picks actions $u_{t}$ and $v_{t}$};
    \node[block, below of = state, node distance = 2 cm] (ADL) {\footnotesize Buys $|v_{t}|$ units to meet the ADL-demand};
    \node[block, below of = ADL, node distance = 2.5 cm] (pass) {\footnotesize Unscheduled ADL jobs are passed on to the next time period $t+1$};

    \node [decision, right of = ADL, node distance = 3 cm] (dec) { \footnotesize If $u_{t} \geq 0$};
    \node [block, right of = dec] (sell) {\footnotesize Sells the power to the neighboring microgrids};
  %  \node [block, right of = sell] (exit1) {Exit};
    \node[block, below of = dec, node distance = 2.5 cm] (non-adl) {\footnotesize  Buys $|u_{t}|$ units to meet the non-ADL demand.};

   % \node[decision, below of = dec] (newdec){\footnotesize If \newline $u_{t} \leq 0$};
    \node [block, right of = non-adl, fill=red!40] (exit) {\footnotesize End};
    \node[decision, below of = non-adl, text width =  6.5 em] (batterydec) {\footnotesize If \newline $nd_{t}-u_{t} \geq 0$};
      \node[block, right of = batterydec, text width = 5 em] (fill) {\footnotesize Stores the remaining power in the battery};
 \node [block, left of = batterydec, fill=red!40, text width = 2 cm] (exit2) {\footnotesize End};
 
  %\path [line] (RL) -- (state);
   %\path [line] (state) -- (action);
   \path [line] (state) -- (ADL);
   \path [line] (ADL) -- (pass);
  \path [line] (pass) -- (exit2);
  \path [line] (state) -| (dec);
  \path [line] (dec) -- node {\footnotesize yes}  (sell);
 \path [line] (sell) -- (exit);
  \path [line] (dec) --  node {\footnotesize no} (non-adl);
 % \path [line] (newdec) --  node {\footnotesize no} (exit);
 % \path [line] (newdec) -| node [near end] {\footnotesize yes} (non-adl);
   \path [line] (non-adl) -- (batterydec);
 \path[line] (batterydec) -- node {\footnotesize yes} (fill);
 %\path [line] (batterydec)  -- node {no}(exit);
  \path [line] (fill) -- (exit);
  \path [line] (batterydec) -- node {\footnotesize no}(exit2);

 \end{tikzpicture}
   \caption{Actions of an RL agent at each time instant $t$}
   \label{flow}
 
 \end{figure}
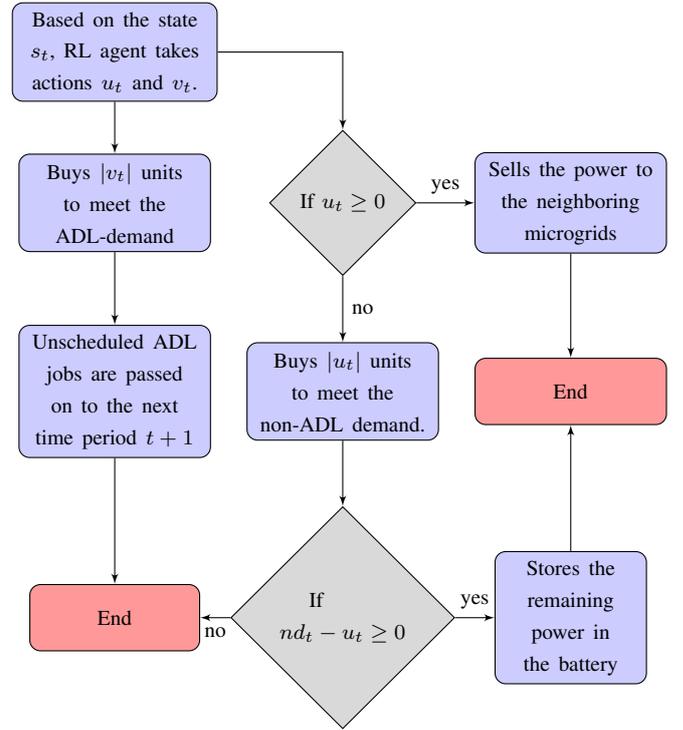

%***************************************************************

\subsection{MDP framework}

MDP is a general framework for modeling problems of dynamic optimal decision making under uncertainty. An MDP is a tuple $<\mathcal{S},\mathcal{U},\mathcal{R},\mathcal{P}>$, where $\mathcal{S}$ is the set of all states, $\mathcal{U}$ is the set of  feasible actions, $\mathcal{R}:\mathcal{S}\times\mathcal{U}\times\mathcal{S}\to \mathbb{R} $ is the single- stage reward function and $\mathcal{P}$ is the state transition probability matrix. In RL, an agent interacts with the environment by observing state $s_t \in \mathcal{S}$ and picking an action $u_t \in \mathcal{U}$. The new state $s_{t+1}$ is obtained from the state transition probability $  \mathcal{P}(s_{t+1} | s_t,u_t)$ and yields a reward $g_{t} = \mathcal{R}(s_t,u_t,s_{t+1})$. The goal of the RL agent is to learn the optimal sequence $\{u_t\}$ of actions so as to maximize its average expected return (see Section \ref{subsec:avg}).

% by selecting actions $u_t$ at every time instant. We assume that actions are taken from the policy $\pi : \mathcal{S} \times \mathcal{U} \to [0,1]$. $\pi(u_t | s_t)$ denotes the probability of selecting action $u_t$ when agent is in state $s_t$. 
%We modeled our problem in the framework of MDP. 
We begin by specifying the states, actions and single-stage rewards, for our MDP model.
%We do not consider any network losses in the proposed model. However, it can be easily integrated in our MDP model. 
\subsubsection{State space}
The state $s_{t}^{i}$ at time instant $t$  for the microgrid $i$ is the following tuple:
\begin{align} \label{stateinfo}
s_{t}^{i} = (t,nd_{t}^{i},p_{t}, J_{t}^{i}),
\end{align}
where the net demand $nd_{t}^{i} = r_{t}^{i} + b_{t}^{i} - d_{t}^{i}$.
%, signifies excess power or deficit depending on whether $nd_{t}^{i} > 0$ or $nd_{t}^{i} < 0$.
If $nd_{t}^{i} > 0$, then  there is excess of power after meeting the non-ADL demand and if $nd_{t}^{i} < 0$, there is a deficit in power even to meet the non-ADL demand. The state also includes time $t$ since optimal action can depend on it. For example, a microgrid operating on solar renewable generation can sell excess power during the morning as the solar power will be available even during afternoon. But it may not be a good choice to aggressively sell it in the evening as there will be no solar power generation during the night. 
\subsubsection{Action space}
Let $P_{t}^{i}$ be the power set of $J_{t}^{i}$, which consists of all possible combinations of the ADL jobs that can be scheduled at time instant $t$ at microgrid $i$. We define another set $A_{t}^{i}$, that denotes the total aggregated ADL demand for each element in $P_{t}^{i}$. For example, the $j^{th}$ element $A_{t}^{i}(j) = \sum_{k=1, \gamma_k^i \in P_{t}^{i}(j) }^n a_k^i$, where $ P_{t}^{i}(j)$ is the $j^{th}$ element in  $P_{t}^{i}$ and $n$ is the total number of elements in $P_{t}^{i}(j)$.

At each time instant $t$, the microgrid controller needs to make two decisions $u_{t}^{i}$ and $v_{t}^{i}$.
\begin{itemize}
\item When $u_{t}^{i}$ is negative, $|u_{t}^{i}|$ represents the amount of power drawn from the peer microgrids/maingrid to meet the non-ADL demand along with optionally storing in the battery (if the power is bought, it is first used to meet the non-ADL demand).
\item When $u_{t}^{i}$ is positive, $|u_{t}^{i}|$ represents the amount of power sold to the peer microgrids/maingrid from the battery storage and energy generated from renewable sources. 
\item The second action $v_{t}^{i}$ pertains to the scheduling decision of ADL jobs taken by microgrid controller $C_i$. $v_{t}^{i}$ is always non-positive, and $|v_{t}^{i}|$ represents the power needed to meet the ADL demand in time interval $t$ at microgrid $i$. Formally, if the $j^{th}$ element of the set $P_{t}^{i}$ is selected at time $t$, $|v_{t}^{i}|$ is equal to $A_{t}^{i}(j)$.
%It denotes the power units needed to satisfy the set of ADL jobs selected in time interval $t$ at microgrid $i$.
\end{itemize}
%The first action $u_{t}^{i}$, if positive, denotes the number of units that the microgrid is willing to sell and if negative, represents the number of units that the microgrid is willing to buy.
 
%Subsequently, a decision $v_{t}^{i}$ is made that denotes the feasible ADL-set that can be scheduling using remaining power in $u_{t}^{i}$. Formally, if $j^{th}$ element of $P_{t}^{i}$ is selected at time $t$ by microgrid $i$, then $v_{t}^{i}$ is equal to $A_{t}^{i}(j)$. 
%Finally the remaining power (if any) is used to fill the battery.
The feasible region for the action $u_{t}^{i}$ is as follows:% Formally, if $j^{th}$ element of $P_{t}^{i}$ is selected at time $t$ by microgrid $i$, then $v_{t}^{i}$ is equal to $A_{t}^{i}(j)$. 
%The second action $v_{t}^{i}$ pertains to the scheduling decision of ADL jobs taken by microgrid $i$. 
%
%It denotes the power units needed to satisfy the set of ADL jobs selected at time $t$ at microgrid $i$. Note that $v_{t}^{i}$ is always non-positive.  
%Let $P_{t}^{i} = \{\Gamma_{1}^{i},\ldots,\Gamma_{N}^{i}\}$ be the power set of $J_{t}^{i}$, which consists of all possible combinations of the ADL jobs that can be sheduled at time instant $t$ at microgrid $i$. 
%Let  $A_{t}^{i} = \{A(\Gamma_{1}^{i}),\ldots,A(\Gamma_{N}^{i})\} $, where $A(\Gamma_{j}^{i}) = \sum_{\gamma_{k}^{i} \in \Gamma_{j}^{i} } a_{k}^{i}$.

\begin{align}
-min(M, B_{i} - nd_t^i + &\max_{1\leq j \leq 2^n} A_t^i(j) ) \leq u_t^i + v_t^i \nonumber\\ &\leq max(0, nd_t^i),
\end{align}

where $M$ denotes the maximum amount of power a microgrid can buy to meet the demand. This constraint is to maintain the stability of the main grid. The above bounds indicate that the  microgrid can sell the surplus; or can buy energy to meet the non-ADL demand, ADL demand and to fill its battery. Note that there is flexibility for microgrids to buy (sell) this power from (to) the neighboring microgrids. If it needs to buy (sell) more power, only then it buys (sells) it from (to) the main grid. In this way, we allow for cooperation among the microgrids. Further this makes sure that the demand at the main grid is controlled at all the times. 
%The intuition behind the bounds is as follows. A microgrid can sell at most the excess power. That is, the power remaining after meeting the demand. While buying, it can buy to meet the demand (both ADL and non-ADL) and also to fill its battery.

%After the controller picks action $u_{t}^{i}$, we construct the feasible set $F_{t}^{i}$ (an element of $P_{t}^{i}$) that can be scheduled with $u_{t}^{i}$. More formally, each element $j$ of  $F_{t}^{i}$ has to satisfy the following condition:  $A_t^i(j) \leq u_{t}^{i} $, where $A_t^i(j)$ is the total energy required to finish all the ADL jobs in it. The controller picks action $v_{t}^{i}$ which is an element of $F_{t}^{i}$, which results in scheduling all the ADL jobs in that subset. The remaining power is used to meet the non-ADL demand or for storage in the battery.

%Each element $\Gamma_{j}^{i}$ in the $F_{t}^{i}$ has to satisfy the following condition $A(\Gamma_{j}^{i}) \leq u_{t}^{i} $.
%Agent has to pick the action $v_{t}^{i} = \Gamma_{j}^{i} \in F_{t}^{i}$. Now ADL jobs in $\Gamma_{j}^{i}$ will get sheduled. 
 Let $\widehat J_{t+1}^{i}$ be the new set of ADL jobs received by the microgrid $i$ in the time interval $t+1$. Depending on the action $v_{t}^{i}$, not all the ADL jobs might have got scheduled in the time interval $t$. The ADL jobs which are not scheduled in the time interval $t$ but are eligible to be scheduled beyond time interval $t$ are considered in the time interval $t+1$ along with the new jobs $\widehat J_{t+1}^{i}$. Thus, we have $J_{t+1}^{i} = \widehat J_{t+1}^{i} \cup \widetilde J_{t}^{i}$, where $\widetilde J_{t}^{i} =  \{\,(a_{j}^{i}, f_{j}^{i}-1) \,\,|\,\, (a_{j}^{i}, f_{j}^{i}) \in \overline J_{t}^{i} \,\, and \,\, f_{j}^{i} > 0 \, \}$ and $\overline J_{t}^{i} = J_{t}^{i} - P_{t}^{i}$.
 
%  \st{
%  Let $\widehat J_{t+1}^{i}$ be the new set of ADL jobs received by the microgrid $i$ in the time interval $t+1$. Depending on action $v_{t}^{i}$, some of the ADL jobs will not get scheduled. These are then considered in time step $t+1$, if they can be scheduled without incurring any penalty. The set of all ADL jobs at time instant $t+1$ is then the union of the new and old ADL jobs, where the latter are those that are not scheduled even after reducing $f_{j}^{i}$ by one (number of future time slots remaining by which one can schedule that job without incurring penalty). Thus, we have $J_{t+1}^{i} = \widehat J_{t+1}^{i} \cup \widetilde J_{t}^{i}$, where $\overline J_{t}^{i} = J_{t}^{i} - P_{t}^{i}$ and $\widetilde J_{t}^{i} =  \{\,(a_{j}^{i}, f_{j}^{i}-1) \,\,|\,\, (a_{j}^{i}, f_{j}^{i}) \in \overline J_{t}^{i}\}$.
%  }
%For the next time instant $t+1$, we update the following :

The battery information is updated as follows:
\begin{align}
b_{t+1}^{i} = max(0,nd_{t}^{i} - u_{t}^{i}),
\end{align}
which denotes the power available after meeting the non-ADL demand.
Figure \ref{flow} illustrates the actions of a microgrid at every time instant $t$.
\subsubsection{Single-stage reward function}
We want to maximize the profit of each microgrid obtained by selling power while reducing the demand and supply deficit. Our single-stage reward function has components for both the reward obtained by selling power and penalty for unmet demand. The single-stage reward  function for the microgrid $i$ at time $t$ is as follows:
\begin{align}\label{reward}
g^{i}(s_t^i,u_t^i,v_t^i) =& p_{t}*(u_{t}^{i} + v_{t}^{i}) + c*min(0,nd_{t}^{i} - u_{t}^{i})  \nonumber\\ & - c* \sum_{k =1}^{n} I_{\{f_{k}^{i} = 0\}} a_{k}^{i} .
\end{align}
The first term in \eqref{reward} represents the loss/gain incurred for  buying/selling  power while the second and third terms represent the penalty  incurred for not meeting the non-ADL and the ADL demands respectively. Here, $c \,\, (\ge 0)$ is the penalty per unit of unmet demand and $I_{\{f_{k}^{i} = 0\}}$ is the indicator random variable which equals one if $f_{k}^{i} =0$ and is zero otherwise. Here $c$ acts as a threshold between the profit of the microgrid and the penalty for not satisfying the demand. For example, when $c = 0$, each microgrid takes decision to maximize its profit without satisfying any customer demand. On the other hand, if the value of $c$ is very high, microgrids need to satisfy customer demand at every time instant as they incur huge penalty otherwise.
%The microgrid incurs a cost of $c$ for every unit of demand that is not met. 
%******Need to write about transition probability kernel ***********.
Next, we provide the long-run average cost objective function. 
\subsection{Average cost setting} \label{subsec:avg}
The objective is to maximize the expected average profit obtained by all the microgrids. 
The long-run average profit objective function $J^{i}(\pi)$ of the microgrid $i$ for a given policy $\pi$ is given as follows:
%\cite{avgcost}:
\begin{align}
J^{i}(\pi) := \lim \sup_{n \rightarrow \infty}\frac{1}{n} \,\, \E \left(\left.\sum_{t = 0}^{n} g^{i} (s_{t},u_{t},v_{t})\right|\pi \right),
\end{align}
where $E(.)$ denotes the expected value. Here we view a policy $\pi$ as the map $\pi : S \to A$ which assigns for any state $s$, a certain feasible action $a$. The goal of our RL agent $i$ is to find $\pi_{i}^* = \argmax_{\pi \in \Pi} J^{i}(\pi)$, where $\Pi$ is the set of all feasible policies.

\section{Algorithm}\label{sec:algo}

%We first note that the renewable generation is uncertain in nature. 
%That is, we do not know in the current time period, the renewable generation in the future time periods.

In this work, we do not assume any model of the system (i.e., probability transition model of the demand, supply as well as renewable energy generation). We apply Reinforcement Learning algorithms that do not assume any model of the environment to provide optimal solution. We assume that we have access to a simulator that provides the state samples (i.e., Non-ADL and ADL demand, price) at every time instant to the algorithm. This can be achieved, for instance, through smart meters deployed in households. We employ the Average-Cost Q-Learning algorithm, a  popular RL method for solving the average cost problem in section \ref{subsec:avg}.
%To solve the above average cost problem, we apply a popular RL algorithm, Q-Learning.
We apply the Relative Value Iteration (RVI) based Q-Learning algorithm for each agent $i$, proposed in \cite{avgcost}. The algorithm is described below.

Let $Q_{t}^{i}(s^{i}_{t},u^{i}_{t})$ represent the Q-value estimate corresponding to state $s^{i}_{t}$ and action $u^{i}_{t}$ for the agent $i$ in the $t^{th}$ iteration. The initial Q-values associated with all states and actions are set to zero i.e., $Q_{0}^{i}(s^{i},u^{i}) = 0$ $\forall \hspace{0.1 cm} (s,u)$ and $\forall i$. Subsequently, the Q-values are updated as follows (using similar notations as in \cite{avgcost}):  
\begin{align} \label{avg_first}
Q^{i}_{t+1}(s^{i}_{t},u^{i}_{t}) &= Q^{i}_{t}(s^{i}_{t},u^{i}_{t}) + \alpha(t,s^{i},u^{i})(g^{i}(s^{i}_{t},u^{i}_{t},s^{i}_{t+1}) + \nonumber\\ &  max_{u} Q^{i}_{t}(s^{i}_{t+1},u) - f(Q^{i}_{t})- Q^{i}_{t}(s^{i}_{t},u^{i}_{t})),
\end{align}

where $\alpha(.)$ is the learning rate, $g^{i}(s^{i}_{t},u^{i}_{t},s^{i}_{t+1})$ is the reward obtained by taking an action $u^{i}_{t}$ in state $s^{i}_{t}$ and transitioning to the state $s^{i}_{t+1}$ and $f(Q^{i}_{t})$ is a Lipschitz continuous function satisfying suitable conditions specified in \cite{avgcost}. This term is subtracted from \eqref{avg_first} to maintain stability of the update equation. It is shown in \cite{avgcost} that $f(Q^{i}_{t})$ converges to the optimal average cost as $t \rightarrow \infty$. One such function $f(.)$ that satisfies the conditions is $max_{u} Q^{i}_{t}(s^{'},u)$, where $s^{'}$ is an arbitrarily chosen, fixed state. Therefore the final update equation for the Q-values for each agent $i$ is as follows: 

\begin{align} \label{avg_second}
Q^{i}_{t+1}(s^{i}_{t},u^{i}_{t}) &= Q^{i}_{t+1}(s^{i}_{t},u^{i}_{t}) + \alpha(t,s^{i},u^{i})(g^{i}(s^{i}_{t},u^{i}_{t},s^{i}_{t+1}) + \nonumber\\ &  max_{u} Q^{i}_{t}(s^{i}_{t+1},u) - max_{u} Q^{i}_{t}(s^{'},u) - Q^{i}_{t}(s^{i}_{t},u^{i}_{t})),
\end{align}

% \begin{align} \label{avg_second}
% Q^{n+1}(s,u) &= Q^{n}(s,u) + \alpha(n,s,u)(g(s,u,s^{'}) + \nonumber\\ &  max_{u} Q^{n}(s^{'},u) - max_{a} Q^{n}(s_{0},a) - Q^{n}(s,u)).
% \end{align}

At each iteration of this algorithm, the agent $i$ selects an action $u^{i}_{t}$ for the current state $s^{i}_{t}$ using the $\epsilon-$greedy policy. That is, a random action is selected with probability $\epsilon$ and the action that maximizes the current Q-estimate is selected with probability $1-\epsilon$. The simulator takes the current $(s^{i}_{t},u^{i}_{t})$ pair and generates the current stage reward and the next state. In \cite{avgcost}, it is shown that under an appropriate learning rate, the algorithm converges to the optimal Q-values which implicitly give the optimal policy. 
Each microgrid runs a version of this algorithm independently until convergence. 
%This process is continued till convergence, at which stage, we can select the optimal action for the state $s$ to be the action $u$ that maximizes the Q-value $Q^{i}_{*}(s,u)$. 
The optimal policy of microgrid $i$ is obtained as follows:
\begin{align}
\pi_{i}^{*}(s) = \argmax_{u}Q_{*}^{i}(s,u)
\end{align}
that is, the optimal action in state $s$ is obtained by taking the maximum over all actions of the Q-values in state $s$.  

%The optimal policy of microgrid $i$ is obtained as follows:
%\begin{align}
%\pi_{i}^{*}(s) = max_{u}Q^{i}(s,u),
%\end{align}
%that is, the optimal action in state $s$ is obtained by taking the maximum over all actions of the Q-values in state $s$.  

\section{Simulation Experiments}\label{sec:experiments}
%We used the RAPsim  simulator  \cite{rapsim} for the evaluation of our algorithms. RAPsim allows users to simulate microgrid networks (involving main grid, microgrids with solar and wind power generating capabilities, individual homes having solar panels). We implement our models on a network with three microgrids (see Figure~\ref{exp}), out of which two operate on solar and one (in the middle) on wind power. The solar microgrid on the right has more generation capacity than the one on the left. Each microgrid can serve power to homes that are only connected to it. 

We implement our model on networks with three and five microgrids, respectively. In the three microgrid setup, two of the microgrids have solar renewable energy as the power supply source while the third operates on wind energy. In the five microgrid network, two of the microgrids use solar renewable energy as the power supply source, two of them use wind renewable energy as the power supply source and one does not have any access to renewable energy. To simulate the renewable generation, we use the RAPsim software \cite{rapsim}. RAPsim is an open source simulator for analyzing the power flow in microgrids. It has a provision for simulating the renewable generation, which is the main feature that we use in our experiments. %We construct our microgrid model as shown in Fig 2. We can see that there are three microgrids, two of them operating on the solar energy and the other on the wind energy. The solar microgrid in the right has more capacity than that of the one in the left. These microgrids also have electrical connections from the main grid. Each microgrid provides power to the respective houses on their power line. 

%\begin{figure}[thbp]
%	\centering
%	\includegraphics [scale = 0.6]{experimental_setup.jpg}
%        \caption{Experimental Setup}
%	\label{exp}
%\end{figure}
\subsection{Implementation}
We solve the MDP model described in section \ref{sec:model} and refer to it as the \textbf{ADL-sharing model}. For comparison purposes, we also solve the following MDP models.\footnote{The implementations of all the three MDP models are available at \newline \url{https://github.com/raghudiddigi/SmartGrid} } 
\begin{itemize}
	\item \textbf{Greedy-ADL model}: In this model, microgrids exhibit greedy behavior. They share power only after filling their respective batteries fully. The actions $u_t^i$ and $v_{t}^{i}$ at each time instant $t$ are bounded as follows:  
	\begin{align}
	-min(M, & B_i - nd_t^i + \max_{1\leq j \leq 2^n} A_t^i(j) ) \leq u_t^i + v_t^i \nonumber\\ &\leq max(0, nd_t^i - B_i).
	\end{align}
	Thus, if $ nd_t^i < 0$, decision is taken on the amount of power to buy in order to satisfy the demand and to fill the battery. If $ nd_t^i > 0$, then the generated excess power by the microgrid $i$ in time interval $t$ is first used to fill the battery fully and if more power is left, it will be sold to the other microgrids/main grid.
	
	\item \textbf{Non-ADL model}:  In this model, ADL demand is treated as normal demand. Penalty is levied immediately if the demand is not met in the current time slot.
%This model is similar to the $ADL-sharing$ model, but without the concept of ADL jobs. In this model, the ADL demand is included in the main demand. Unlike the $ADL-sharing$ model, there is no flexibility of intelligently scheduling the ADL jobs. 	
\end{itemize}
\subsection{Simulation setup}
We used the RAPsim simulator to generate per hour renewable energy data for each of the microgrids. We used this data to fit a Poisson distribution for energy generation at each microgrid. We limit the maximum renewable energy available at each time instant to 8 units. The number of decision time intervals in a day is taken to be 4. %The mean of the Poisson distribution for the two solar and one wind power obtained are as follows :

%$$ \left[ \begin{array}{ccccc}
%	0 & 0.5410 & 6.5965 & 4.3712 \\
%	0 & 0.7350 & 8.6901 & 5.7239 \\
%	3.6087 & 3.2167 & 3.1405 & 3.8590
%\end{array} \right],$$

%where the element $(i,j)$ represents the Poisson mean of microgrid $i$ at time $j$.
For each time period, non-ADL demand ($d_t^i$) at each of the microgrids can be one of the following  three values: 2, 4 or 6 units. The price ($p_t^i$) per unit energy value (in USD) is considered to be one of 5, 10 or 15. 
% For demand at each microgrid, we considered three values - 2, 4 and 6 units. 
%We simulate the above setup for the month of September 2017 in the RAPsim and collect the wind and solar renewable power generated each day every hour. Using this data, we fit Poisson distribution and obtain the Poisson mean. 
%The parameters for our experiments are described below. The number of decision time periods is taken to be 4 (i.e., t = 4). We consider 3 demand values for all the microgrids - 2, 4 and 6 units. 
The transition probability matrix for non-ADL demand and the price values are generated randomly.

%\[P_{1}= \left[ \begin{array}{ccc}
%0.2 & 0.6 & 0.2 \\
%0.1 & 0.2 & 0.7 \\
%0.8 & 0.1 & 0.1
%\end{array} \right],
%%
%%P_{2}=
%\left[ \begin{array}{ccc}
%0.2 & 0.2 & 0.6 \\
%0.8 & 0.1 & 0.1 \\
%0.2 & 0.7 & 0.1
%\end{array} \right]
%\]
%
%\[P_{3}= \left[ \begin{array}{ccc}
%0.5 & 0.5 & 0 \\
%0 & 0.5 & 0.5 \\
%1 & 0 & 0
%\end{array} \right],
%%
%Q=
%\left[ \begin{array}{ccc}
%0.2 & 0.4 & 0.4 \\
%0.1 & 0.5 & 0.4 \\
%0.5 & 0.4 & 0.1
%\end{array} \right]
%\]
For our experiments, the maximum size of the battery ($B_i$) is set to 8 units and the maximum power that a microgrid can obtain from the main grid ($M$) is limited to 14 units.
At each microgrid, we consider 3 ADL jobs, $\{\gamma_{1}^{i} =  (1,2), \gamma_{2}^{i} =  (1,3),  \gamma_{3}^{i} =  (2,4)\}$ at the start of the day, where ADL job $\gamma_{j}^{i} =  (a,b)$ requires $a$ units of demand within $b \,\,(> 0)$ time slots. In the $Non-ADL$ model, the ADL demand is added to the demand at $t = 1$ each day. %We considered each day is having three time slots. 
We ran all our simulations for each of the following $c$ (penalty in USD per unit of unmet demand) values : 0, 5, 10 and 30, respectively.
\subsection{Results}
The algorithms are trained for $10^7$ cycles. We used the average profit obtained by each microgrid as a performance metric to evaluate the models. 

\begin{figure}[thbp]
	\centering
	\includegraphics[scale = 0.2]{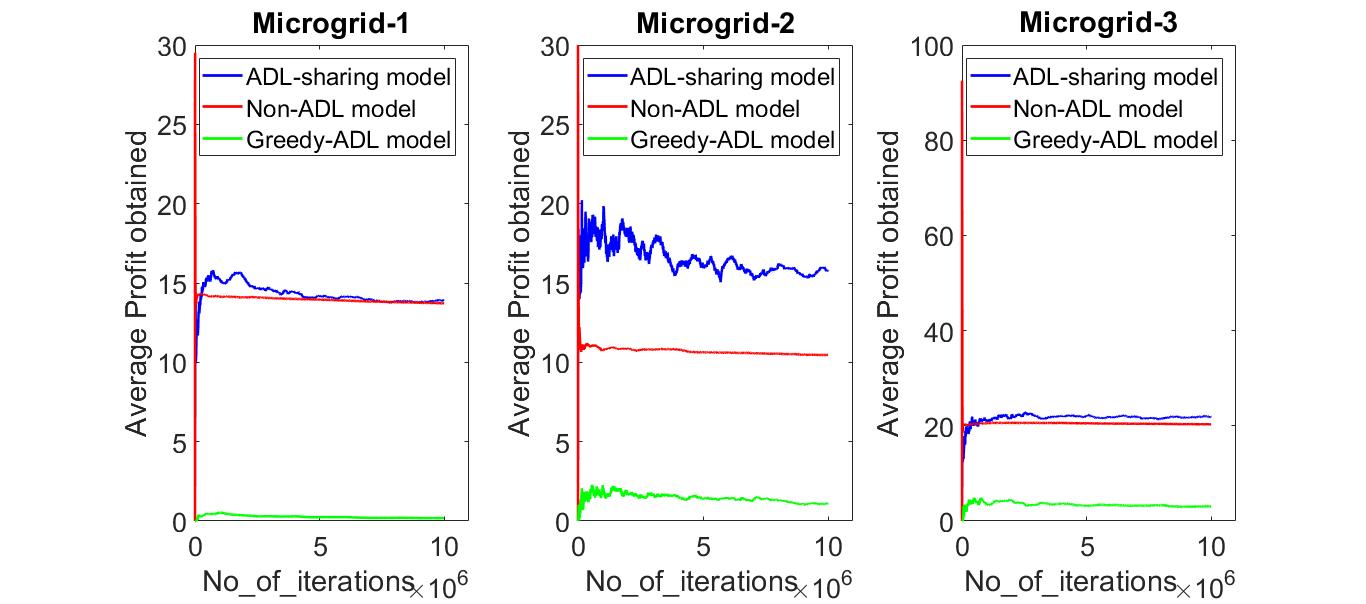}
	\caption{Convergence of algorithms for the three models when $c = 0$ for the three microgrid network as a function of number of iterations.}
        \label{t1}
\end{figure}

\begin{figure}[thbp]
	\centering
    \begin{subfigure}[b]{\textwidth}
    	\includegraphics[scale = 0.2]{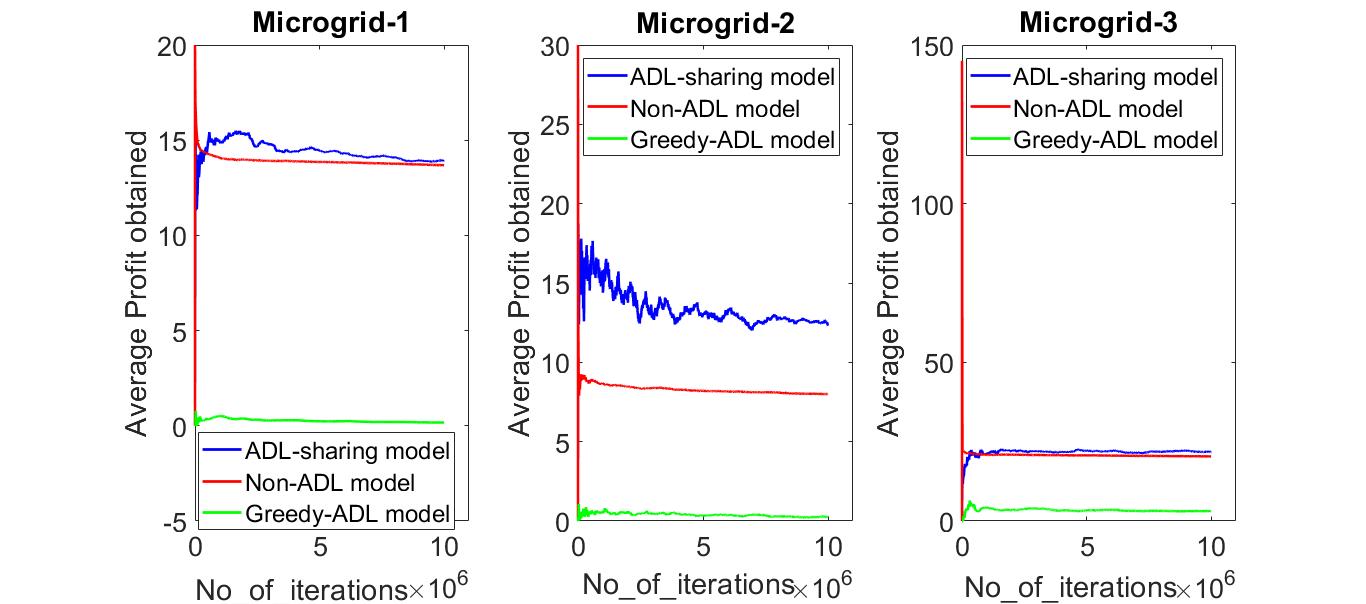}
    \end{subfigure}
    \begin{subfigure}[b]{\textwidth}	
	\includegraphics[scale = 0.2]{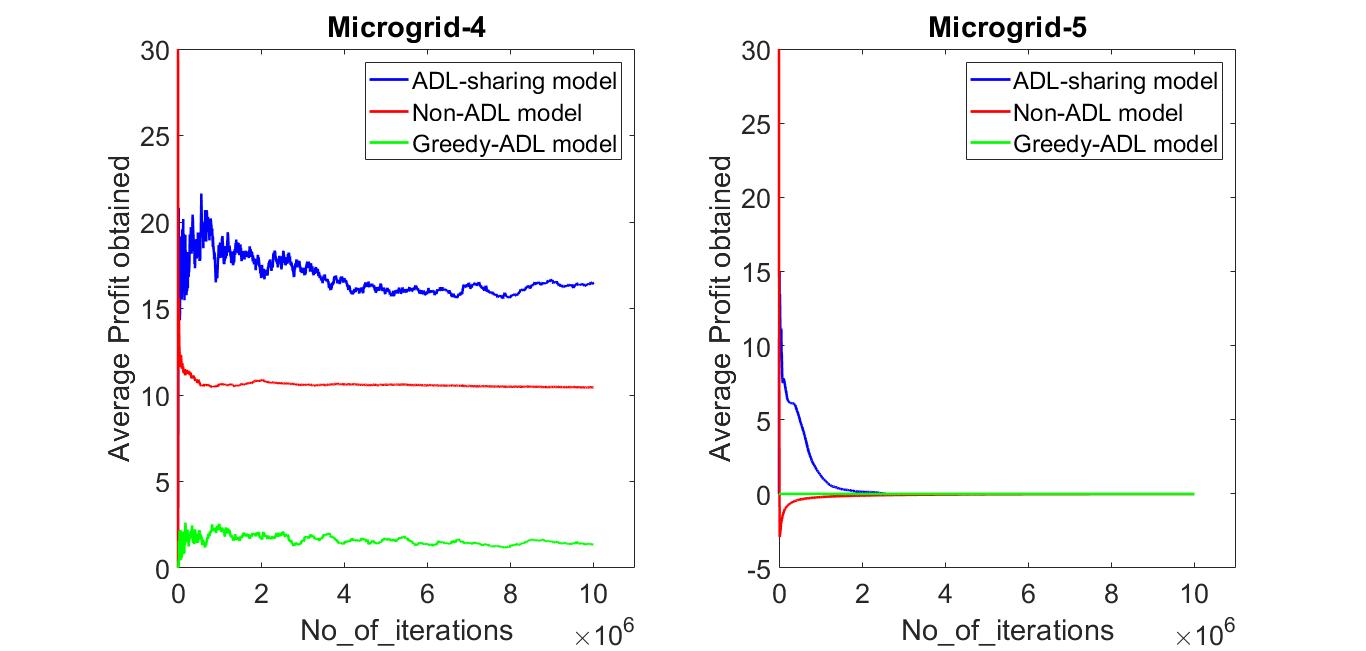}
    \end{subfigure}
    
	\caption{Convergence of algorithms for the three models when $c = 0$ for the five microgrid network as a function of number of iterations.}
        \label{f1}
\end{figure}

Figures ~\ref{t1} and ~\ref{f1} plot the average profit obtained for each microgrid in the two settings of three and five microgrid networks, respectively, versus the number of iterations, when $c = 0$ in each case. We can see that the algorithms show convergence as the number of iterations increase.

\begin{figure}[thbp]
	\centering
	\includegraphics[scale = 0.2]{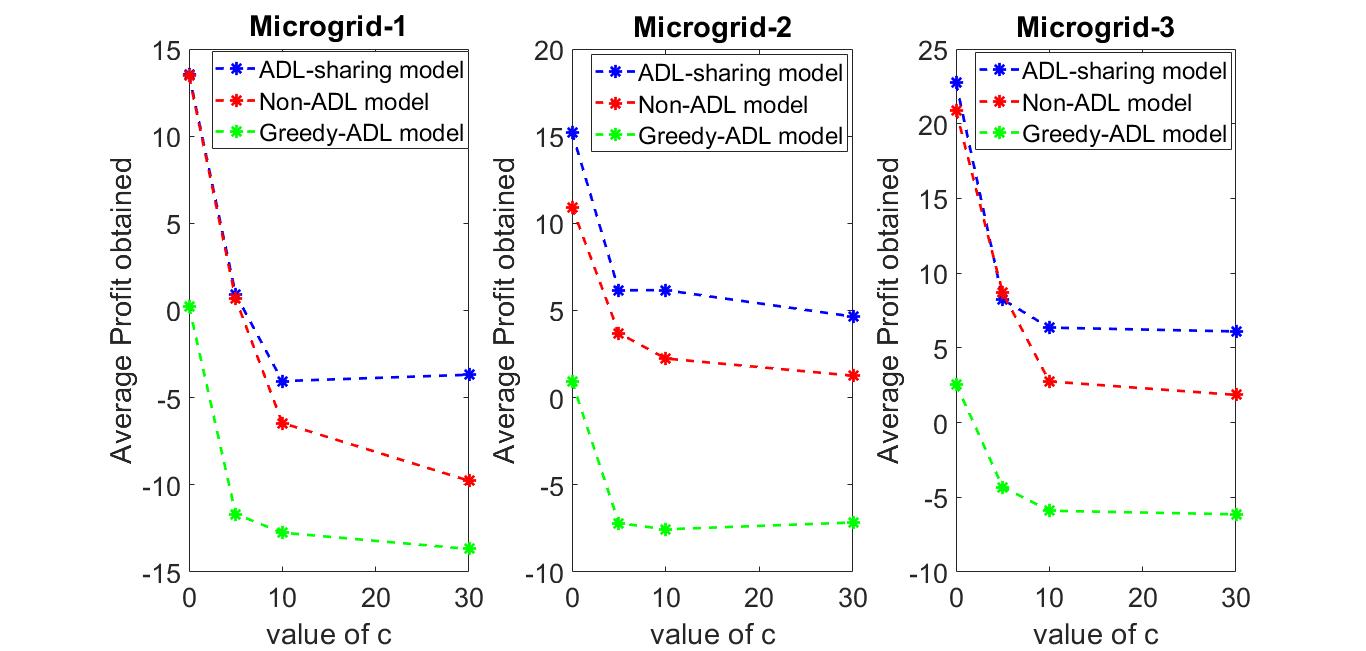}
	\caption{Performance Comparisons of the three models on each microgrid in the three microgrid network as a function of $c$.}
        \label{t2}
\end{figure}

\begin{figure}[thbp]
	\centering
    \begin{subfigure}[b]{\textwidth}
    	\includegraphics[scale = 0.2]{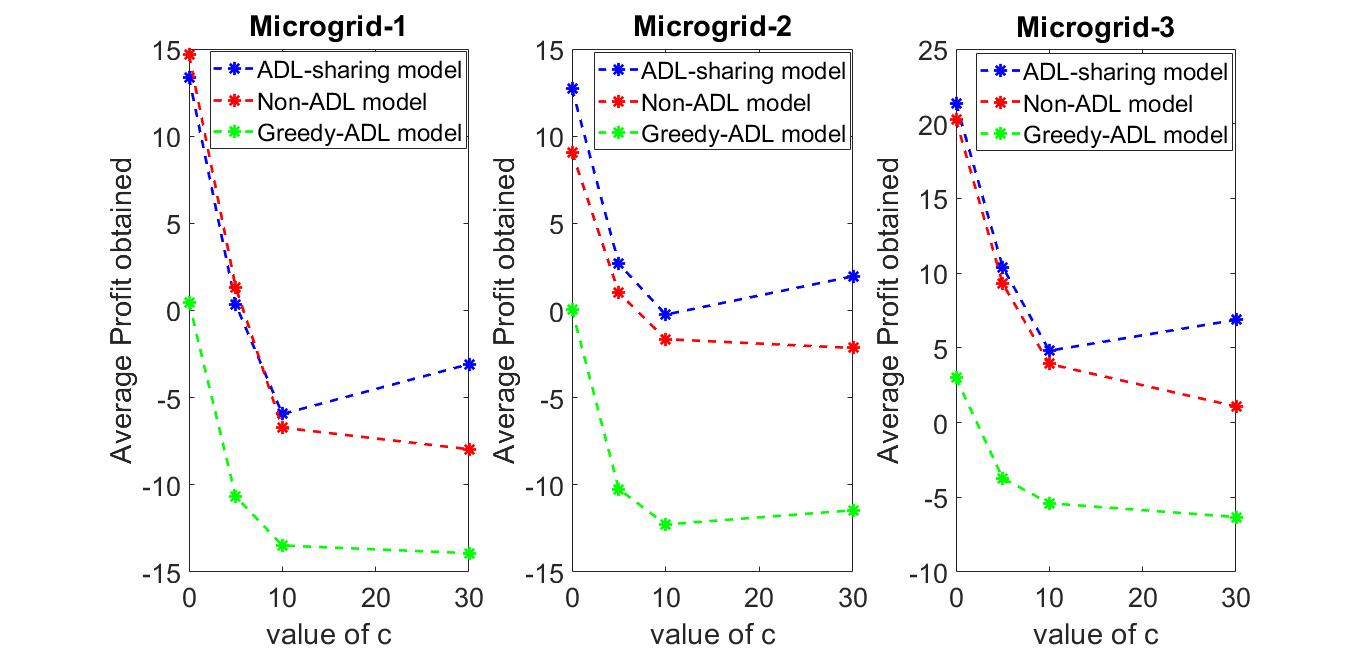}
    \end{subfigure}
    \begin{subfigure}[b]{\textwidth}	
	\includegraphics[scale = 0.2]{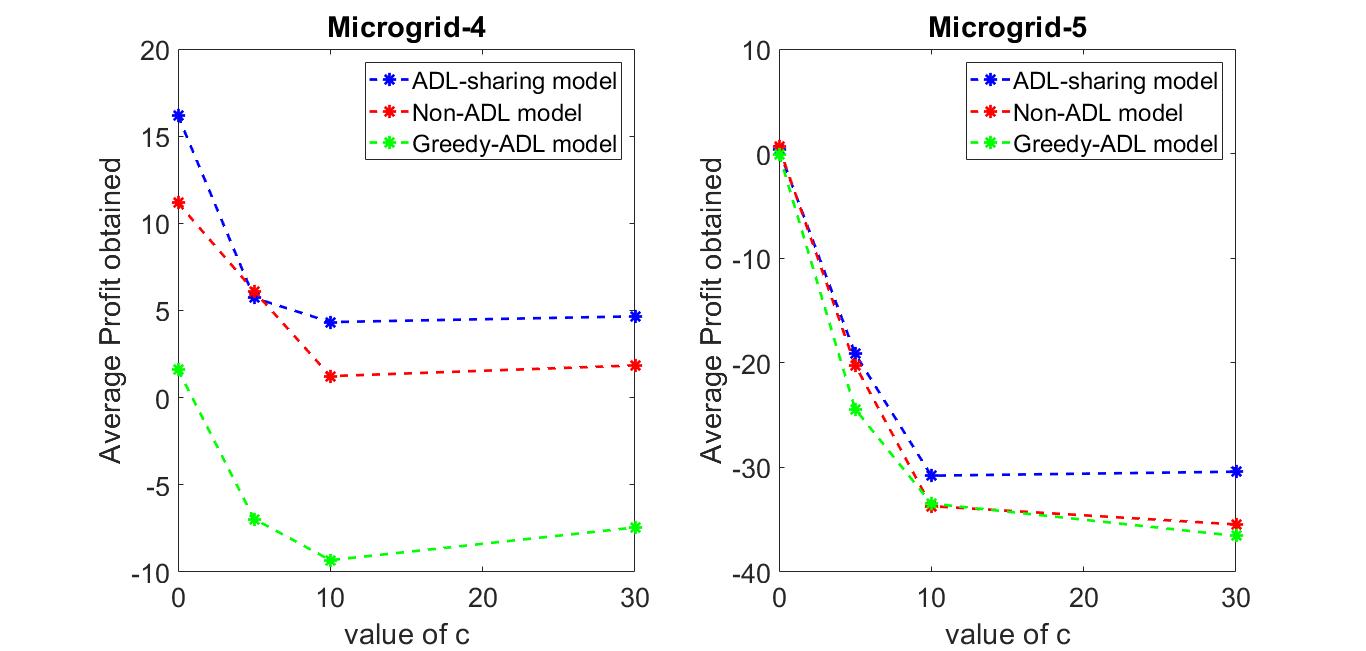}
    \end{subfigure}
    
	\caption{Performance Comparisons of the three models on each microgrid in the five microgrid network as a function of $c$.}
        \label{f2}
\end{figure}

Figures ~\ref{t2} and ~\ref{f2} on the other hand plot the average profit obtained for each microgrid versus $c$ (i.e., penalty per each unit of unmet demand) for all the three models. We run the trained models for 1000 runs to obtain the average profit in each case.  

\subsection{Discussion}

%\begin{itemize}
	
%\item In Figure~\ref{r2}, we observe that as the number of iterations increase, the average profit increases till the convergence.

%\item In  the $Greedy-ADL$ model there is less buying and selling of power compared to the other models.
%In the $Greedy-ADL$ model the sharing of power is done only after filling the battery. Therefore, even though there is less buying of power in $Greedy-ADL$, there is very little selling of power as well.
% Therefore the overall profit obtained is not high. Thus intelligent sharing of power among microgrids as with the RL technique, yields more profit than in the $Greedy-ADL$ case. 

%\item We also observe that, $ADL-sharing$ model outperforms the $Non-ADL$ model. In $ADL-sharing$, there is a flexibility to intelligently schedule the ADL jobs according to the non-ADL demand and price. 
%But in the $Non-ADL$ case, the penalty is immediately levied if the demand (including the ADL demand) is not met. 
%Hence we conclude that intelligently scheduling the ADL demand results in better performance.

From our experiments we make the following observations:
\begin{itemize}
\item When $c = 0$, the microgrid controllers need not buy power to satisfy the excess demand as they do not incur penalty for not meeting the demand. In the $ADL-sharing$ and $Non-ADL$ models, we observe that all the controllers fill the battery when the price is low while they sell power when the price is high. %Hence there will be not much sharing among the agents. 
Hence, the profit obtained is very high compared to the $Greedy-ADL$ model where the power is bought to first fill the battery.

\item As the value of $c$ increases, we expect that the profit earned by the microgrid would decrease. This is because the penalty for not meeting the demand would increase. From Figure ~\ref{f2}, we observe that profit for $c = 30$ is slightly higher than when compared with the case of $c = 10$ for $ADL-Sharing$ model. This is due to the high variance in the solar generation during the testing phase. When the value of $c$ is set to 30, the demand at all times will be met without any penalty. %Therefore, this case is more resistant to the variance in renewable generation than the case when $c= 10$. 

\item We observe that the profit gap between the $ADL-sharing$ method and the $Non-ADL$ method increases as the value of $c$ increases. This shows that the profit of the microgrids increases due to the flexibility available to the controllers in scheduling the demand by following the $ADL-sharing$ model. 

\item The sharing among the microgrids happens as follows: A microgrid operating on solar renewable energy source in the second time period shares the excess power with the other microgrids, as it generates more power as the day progresses. At the same time, a microgrid operating on wind renewable energy source buys power to store in its battery, if it expects more demand than the power it generates in future time periods.

\end{itemize}

From the above discussion we conclude that our proposed $ADL-sharing$ model provides more profits by exhibiting the following intelligent behavior: 
\begin{enumerate}[label=(\alph*)]
\item Schedules few of the ADL jobs at the beginning, few at the end and few in the middle of their allowed execution time window to exploit flexible nature of the ADL demand. 
\item Microgrids do not sell all of their surplus energy to the other microgrids if there is more demand than supply in the future (particularly, solar microgrids sell excess energy during the midday but not at the end of the day).
\end{enumerate}

\section{Conclusion}\label{sec:conclusion}
Providing a unified solution framework for modeling both demand-side management problem (scheduling ADL jobs) and supply-side management problem (enabling cooperative energy exchange among the microgrids) is a challenging task, particularly when both demand and supply are considered stochastic. We have studied these two problems, for the first time, in a unified framework by using MDPs. Also, for the first time in the literature, we proposed the method of scheduling ADL demand at the microgrid level as a load shifting technique. RL algorithms provide an optimal solution methodology for solving MDP when the underlying model is not known. We apply the Q-learning algorithm to maximize the profit earned by microgrids by selling excess energy while maintaining a low gap between demand and supply. Based on the simulation experiments, we show that the policy obtained by our MDP model (ADL-sharing model) consistently outperforms the policies obtained by other models.

As future work, we would like to consider the pricing mechanism for microgrids. In the current model, the transaction of power is carried out at the price decided by the main grid. The pricing mechanism allows microgrids to bid for the selling price as well as buying price. One can use RL agents to bid for adaptive prices in such a way that microgrids maximize their profits. Another important future work is to use RL algorithms with function approximation to scale the proposed algorithms. The challenge here is to select the appropriate features to obtain an optimal policy.

%\section*{Acknowledgment}

%The authors would like to thank Robert Bosch Centre for Cyber-Physical Systems, IISc, Bangalore, India for supporting part of this work.

 \bibliographystyle{IEEEtran}
 \bibliography{IEEEabrv,reference}

\end{document}